\documentclass[prb,twocolumn,superscriptaddress,showpacs,amsmath,amssymb,floatfix]{revtex4-1}
\usepackage{amsfonts}
\usepackage{bm}
\usepackage{color}

\usepackage{graphicx}
\usepackage{epstopdf}

\begin{document}
\title{On the superconductivity of graphite interfaces}

\author{Pablo Esquinazi}
\affiliation{Division of Superconductivity and Magnetism, Institut f\"ur Experimentelle Physik II,
\\ Universit\"at Leipzig, Linnestrasse 5, D-04103 Leipzig, Germany}

 \author{Tero T. Heikkil\"a}
\affiliation{Department of Physics and Nanoscience Center, University of Jyv\"askyl\"a,
\\
P.O. Box 35 (YFL), FI-40014 University of Jyv\"askyl\"a, Finland}

 \author{Yury V. Lysogorskiy}
\affiliation{Institute of Physics, Kazan Federal University, Kremlevskaya St. 16a, 420008 Kazan,
Russia}

 \author{Dmitrii A. Tayurskii}
\affiliation{Institute of Physics, Kazan Federal University, Kremlevskaya St. 16a, 420008 Kazan,
Russia}
\affiliation{Centre for Quantum Technologies, Kazan Federal University, Kremlevskaya St. 16a,
420008 Kazan, Russia}

\author{Grigory E.~Volovik}
\affiliation{Low Temperature Laboratory, Aalto University,  P.O. Box 15100, FI-00076 Aalto, Finland}
\affiliation{Landau Institute for Theoretical Physics, acad. Semyonov av., 1a, 142432,
Chernogolovka, Russia}

\date{\today}

\begin{abstract}
We propose an explanation for the appearance of superconductivity
at the interfaces of graphite  with Bernal stacking order. A
network of line defects with flat bands appears at the interfaces
between two slightly twisted graphite structures. Due to the flat
band the probability to find high temperature superconductivity at
these quasi one-dimensional corridors is strongly enhanced. When
the network of superconducting lines is dense it becomes
effectively two-dimensional.  The model provides an explanation
for several reports on the observation of superconductivity up to
room temperature in different oriented graphite samples, graphite
powders as well as graphite-composite samples published in the
past.
\end{abstract}
\maketitle



Highly oriented pyrolytic graphite (HOPG) is known to have quasi
two-dimensional (2D) interfaces \cite{ina00,bar08}.
 Recently, it
was found that such interfaces exhibit  extraordinary properties
that indicate the existence of granular 2D superconductivity
within the interfaces and  up to the room temperature or above
\cite{bal13,sch12,schcar}. Here we discuss a possible origin of
this phenomenon.

\begin{figure}[hbt]
\centerline{\includegraphics[width=1.0\linewidth]{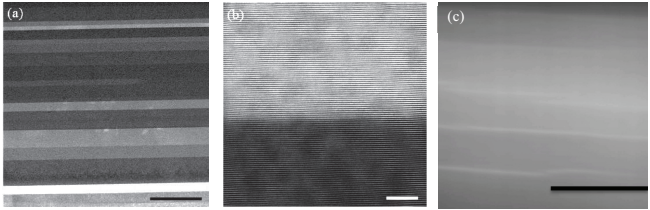}}
\caption{(a) Transmission electron microscope picture of a
HOPG lamella of grade A. The scale bar corresponds to 500~nm. The graphene planes
run parallel to the interfaces. (b) A zoom of a section
of (a) with higher resolution where the  edges of the graphene planes can be recognized.
The $c-$axis is normal to the graphene planes.
The scale bar corresponds to 5~nm.\cite{ab} (c) Similar but for a HOPG sample of grade B and
from a different source. The bar corresponds to $1~\mu$m. In this sample there is less
area with well defined interfaces than in grade A samples.
\label{TEM} }
\end{figure}

The interface in graphite we discuss in this work represents a grain boundary between
domains with slightly different orientations and can be recognised by transmission electron
microscopy with the electron beam applied parallel to the graphene planes of graphite.
Figure~\ref{TEM} shows transmission electron microscopy pictures of
 two HOPG samples at  different resolutions. The interfaces are
at the borders of the crystalline (Bernal stacking order ABA...)
regions characterised by a certain gray colour. The twist angle
$\theta_{\rm twist}$, i.e., a rotation with respect to the
$c-$axis between single crystalline domains of Bernal graphite,
may vary from $\sim 1^\circ$ to $< 60^\circ$, \cite{war09} while
the tilting angle of the grains with respect to the $c$-axis
$\theta_{\rm c} \lesssim  0.4^\circ$ for the highest oriented
pyrolytic graphite samples. When the misfit angle is small enough,
the grain boundary can be represented by a system of dislocations
--  the Burgers-Bragg-Read-Shockley (BBRS) dislocation model
\cite{Burgers1940,Bragg1940,ReadShockley1950}. This is the system
of edge dislocations if $\theta_{\rm c}\neq 0$, and the system of
screw dislocations in the case $\theta_{\rm twist}\neq 0$.

\begin{figure}[hbt]
\centerline{\includegraphics[width=0.90\linewidth]{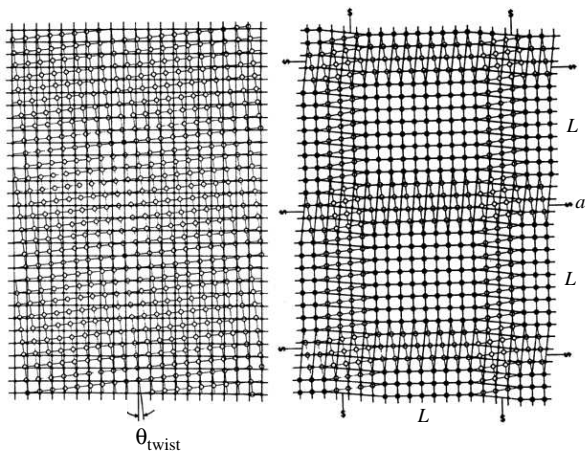}}
\caption{
\label{Fig:Screw}
 The illustration of the dislocation model of the crystal grain boundary \cite{ReadShockley1950}
in the case of the interface between two domains with slightly
different orientations -- a small twist angle $\theta_{\rm
twist}$. On the left is the initial configuration, when two
domains are stuck together; on the right  is the relaxed
configuration of the interface.   The latter consists of perfectly
matched regions of size $L  \sim a/\sin(\theta_{\rm twist}/2)$
separated by the network of linear objects -- the screw
dislocations. For simplicity the interface is illustrated  using
two twisted sheets of quadratic lattice. Here instead of
dislocations, the twist between the layers is mediated by solitons
-- boundaries between the matched regions. }
\end{figure}

The BBRS dislocation model of the interface between two domains with slightly
different orientations --
a small twist angle $\theta_{\rm twist}$ -- is demonstrated in Fig.~\ref{Fig:Screw}.
For simplicity
the interface is illustrated  using two twisted sheets forming square lattices.
In Fig.~\ref{Fig:Screw} ({left}) is the initial configuration, when two domains are stuck together;
in Fig.~\ref{Fig:Screw} ({right})  is the relaxed configuration of the interface.   The latter
consists of perfectly matched regions
separated by the network of the linear objects -- solitons in the case of two sheets and
screw dislocations in the case of real interface.
 The size $L$ of the perfect regions is determined by
$\theta_{\rm twist}$ in the equation  $L \sim a/\sin(\theta_{\rm
twist}/2)$, where $a$ is the interatomic distance.\cite{ReadShockley1950}
For bilayer graphene with slightly
twisted layers, the solitons and their networks can be found in Refs.
\onlinecite{Alden09072013,San-Jose2013,MelePRB2014}.

 The network of linear defects is formed  when the
twist angle is small enough. For bilayer graphene the defects emerge when
$\theta_{\rm twist} \lesssim
1^\circ$. \cite{San-Jose2013} For larger
angles the configuration of the type of Fig. \ref{Fig:Screw}({left})
is preferable, in which the twist angle between the layers
does not change. This configuration gives rise to
Moir\'e patterns as has been reported in the literature recently,
see, e.g., \onlinecite{war09,Bistritzer2011,Brihuega2012,Flores2013,yin14}.

Graphite represents the ordered or disordered array of the two-dimensional graphene sheets. Graphene
is the topological material, which belongs to the class of  topological semimetals, \cite{Burkov2011}.
Its electronic energy spectrum has topologically protected point nodes. \cite{Volovik2007}
In graphite, the point nodes in each layer transform to  the chain of the electron and hole
Fermi surfaces, \cite{McClure1957} which corresponds to approximate line of zeroes protected by topology.
That is why graphite experiences (at least approximately)
the properties, which are generic for the topological matter. \cite{HasanKane2010,Shoucheng2011}

In the topological materials, the topological defects such as
dislocations, quantized vortices, domain walls, solitons, grain
boundaries, etc., frequently  contain exotic gapless branches in
the electronic spectrum. In particular, the networks of solitons
in the twisted bilayer graphene contains topologically protected
helical modes, which is the direct consequence of the twist.
\cite{San-Jose2013} For us it is important that among the gapless
branches  there are Dirac points with quadratic and higher order
touching of branches, and the completely dispersionless  branch
with zero energy -- the flat band. The topologically protected
flat band arises at the zig-zag edge of a graphene sheet;
\cite{Ryu2002} inside Abrikosov vortex in Weyl superconductor;
\cite{Kopnin1991,Volovik1994,Volovik1999}
 at the grain boundary in graphene, which is represented
by the chain of the point edge dislocations. \cite{Feng2012} In
graphite, which is only approximately a topological material, the
flat bands are also approximate. Such flat band arises on the
surface or at the interface of the rhombohedral graphite,
\cite{HeikkilaKopniVolovik2011} where it actually represents the
Dirac point with quadratic spectrum and with extremely large mass.
\cite{kop13}

The situation, which is similar to the interfaces in the Bernal
graphite, is discussed for IV-VI semiconductor heterostructures
consisting of a topological crystalline insulator and a trivial
insulator. \cite{TangFu2014} Due to the lattice mismatch between
insulators, the two-dimensional square array of dislocations with
period of 3-25~nm is
 spontaneously formed at the interface, which leads to a nearly flat band there.
The topological origin of this flat band can be understood in
terms of the pseudo-magnetic field created by strain and the
corresponding Landau levels. Note that a similar pseudo-magnetic
field emerges in the strained graphene \cite{Vozmedioano2013}. All
this suggests that in a similar manner the network of screw
dislocations at the graphite interface  may also lead to exotic
branches with almost the flat spectrum. This is supported by
consideration of  the edge dislocations in graphite. They can be
represented as the edges of the extra layers of the graphene
sheets, which as we know contain  flat bands.

The important consequence of the flattening of the electronic spectrum
is the singular density of states $N({\epsilon})$ at $\epsilon \rightarrow 0$. This
 produces ferromagnetism, superconductivity or another ordered state,
with high transition temperature. In particular, in the presence of the flat band
one has
 $N({\epsilon})\propto \delta(\epsilon)$ and one obtains  the linear dependence $T_c\propto g$ of the
critical temperature on the interaction strength in the Cooper
channel.  \cite{Khodel1990,HeikkilaKopniVolovik2011,kop13,kop11}
The quadratic flattening in 1D systems gives $N({\epsilon})\propto\epsilon^{-1/2}$ and the quadratic
dependence $T_c \propto g^2$ of the critical temperature. \cite{Kopaev1970,KopaevRusinov1987}
This is in clear contrast to the exponential
behavior, $T_c\propto \exp(-1/g)$, in conventional superconductivity.
As pointed out in Ref. \onlinecite{TangFu2014},  in the IV-VI semiconductor multilayers
 the transition temperature is unusually high for these materials,
while the strong anisotropy  of the upper critical field reveals
the two-dimensional character of superconductivity.  The authors
of Ref. \onlinecite{TangFu2014} ascribe that to the flat band
emerging from the misfit dislocation array at the interface
between topological and non-topological insulators. This proposal
coordinates with experiments on  highly oriented pyrolytic
graphite, where the unusually high  transition temperature is
reported and which is associated with the graphite interfaces.
\cite{schcar,bal13}

In the highly oriented graphite the dislocation
network at the interface is dense, with $L\sim 10$ nm.
That is why the superconducting state,
if it is formed in
the 1D ``corridors",  has effectively a two-dimensional
nature with possible flux quantization.
  We note that a misfit in the $c$-axis orientation may
also lead to a similar result, because it would give rise to an
array of straight edge dislocations, each containing a 1D flat
band.

There is another,  non-topological source of the flattening of the
electronic spectrum: it is the effect of electron-electron
interaction \cite{Khodel1990,Dolgopolov2014}. In particular, this
mechanism  is operating in the vicinity of the van Hove
singularity. \cite{Volovik1994,Yudin2014} Note, that experimental
STM and STS studies in bilayer \cite{Brihuega2012,Flores2013} as
well as in multilayer graphene \cite{war09,yin14} demonstrated the
existence  of logarithmic van Hove singularities for  $1^\circ
\lesssim \theta_\mathrm{twist} \lesssim 10^\circ$.  The van Hove
singularity appears on the ``light'' (from STM picture) domains of
twisted bilayer, which corresponds to the Moir\'e pattern.

The model of the interface superconductivity, which we propose,
 may account for several details of different publications reporting
superconducting-like signals up to room temperature
in graphite-based as well as
annealed carbon samples in the last 40 years
\cite{ant74,yakovjltp00,silvaprl,yang01,mol02,fel09,fel14,esqpip}.
In particular one can understand why those signals were difficult
to reproduce, not always stable in time, relatively weak, i.e.
appeared to come from a small amount of superconducting mass, and
very sensitive to the preparation conditions. According to Ref. \onlinecite{San-Jose2013},
the topological protection of
the fermion zero modes leaving in the 1D corridors is weak,
and can be broken
by atomic vacancies or small adsorbates.
That is why future theoretical
work should study the influence of doping (through hydrogen, for
example) at the interfaces.
Local transport measurements of single
interfaces in multilayer graphene samples with different twist
angles are of interest.

\newpage

\section{Acknowledgements} \label{Sec:Acknowledgements}
The ideas drawn in this letter are the result of the discussions
at the Workshop on flat bands and room-temperature
superconductivity, held in Otaniemi, Finland, May 30-31. PE
acknowledges discussions with Roman G. Mints during his stay at
Tel-Aviv University, Ana Ballestar for providing the pictures of
her thesis and Dr. Pipple and Dr. B\"ohlmann for the TEM
measurements. G.E.V. and T.T.H. acknowledge the financial support
by the Academy of Finland through its LTQ CoE grant (project
$\#$250280). Y.V.L. and D.A.T. acknowledge financial support
through Russian Government Program of Competitive Growth of Kazan
Federal University.



\begin{thebibliography}{10}

\bibitem{ina00}
M.~Inagaki, {\em New Carbons: Control of Structure and Functions}.
\newblock Elsevier, 2000.

\bibitem{bar08}
J.~Barzola-Quiquia, J.-L. Yao, P.~R\"odiger, K.~Schindler, and
P.~Esquinazi,
  ``Sample size effects on the transport properties of mesoscopic graphite
  samples,'' {\em phys. stat. sol. (a)}, vol.~205, pp.~2924--2933, 2008.

\bibitem{bal13}
A.~Ballestar, J.~Barzola-Quiquia, T.~Scheike, and P.~Esquinazi,
``Evidence of
  \rm{Josephson-coupled} superconducting regions at the interfaces of highly
  oriented pyrolytic graphite,'' {\em New J. Phys.}, vol.~15, p.~023024, 2013.

\bibitem{sch12}
T.~Scheike, W.~B\"ohlmann, P.~Esquinazi, J.~Barzola-Quiquia,
A.~Ballestar, and
  A.~Setzer, ``Can doping graphite trigger room temperature superconductivity?
  \texttt{E}vidence for granular high-temperature superconductivity in
  water-treated graphite powder,'' {\em Adv. Mater.}, vol.~24, pp.~5826--5831,
  2012.

\bibitem{schcar}
T.~Scheike, P.~Esquinazi, A.~Setzer, and W.~B\"ohlmann, ``Granular
  superconductivity at room temperature in bulk highly oriented pyrolytic
  graphite samples,'' {\em Carbon}, vol.~59, pp.~140--149, 2013.

\bibitem{ab}
 Taken from the Ph.D. Thesis of Ana Ballestar, entitled "Superconductivity at
  Graphite interfaces" defended at the University of Leipzig (2014). The high
  resolution TEM pictures were taken in the ''Max Planck Institute of
  Microstructure Physics'' in Halle (Germany) by Dr. Pipple, the lower
  resolution TEM picture was taken by Dr. W. B\"ohlmann at the University of
  Leipzig.

\bibitem{war09}
J.~H. Warner, M.~H. R\"ommeli, T.~Gemming, B.~B\"uchner, and
G.~A.~D. Briggs,
  ``Direct imaging of rotational stacking faults in few layer graphene,'' {\em
  Nano Letters}, vol.~9, no.~1, pp.~102--106, 2009.

\bibitem{Burgers1940}
J.~Burgers, ``Geometrical considerations concerning the structural
  irregularities to be assumed in a crystal,'' {\em Proceedings of Physical
  Society}, vol.~52, pp.~23--30, 1940.

\bibitem{Bragg1940}
W.~L. Bragg, ``Geometrical considerations concerning the
structural
  irregularities to be assumed in a crystal,'' {\em Proceedings of Physical
  Society}, vol.~52, pp.~105--109, 1940.

\bibitem{ReadShockley1950}
W.~T. Read and W.~Shockley, ``Dislocation models of crystal grain
boundaries,''
  {\em Phys. Rev.}, vol.~78, pp.~275--289, May 1950.

\bibitem{Alden09072013}
J.~S. Alden, A.~W. Tsen, P.~Y. Huang, R.~Hovden, L.~Brown,
J.~Park, D.~A.
  Muller, and P.~L. McEuen, ``Strain solitons and topological defects in
  bilayer graphene,'' {\em Proceedings of the National Academy of Sciences},
  vol.~110, no.~28, pp.~11256--11260, 2013.

\bibitem{San-Jose2013}
P.~San-Jose and E.~Prada, ``Helical networks in twisted bilayer
graphene under
  interlayer bias,'' {\em Phys. Rev. B}, vol.~88, p.~121408(R), 2013.

\bibitem{MelePRB2014}
X.~Gong and E.~J. Mele, ``Stacking textures and singularities in
bilayer
  graphene,'' {\em Phys. Rev. B}, vol.~89, p.~121415, Mar 2014.

\bibitem{Bistritzer2011}
R.~Bistritzer and A.~MacDonald, ``Moir\'e bands in twisted
double-layer
  graphene,'' {\em PNAS}, vol.~108, pp.~12233--12237, 2011.

\bibitem{Brihuega2012}
I.~Brihuega, P.~Mallet, H.~Gonz\'alez-Herrero, G.~T.
de~Laissardi\`ere, M.~M.
  Ugeda, L.~Magaud, J.~M. G\'omez-Rodr\'iguez, F.~Yndur\'ain, and J.-Y.
  Veuillen, ``Unraveling the intrinsic and robust nature of van hove
  singularities in twisted bilayer graphene by scanning tunneling microscopy
  and theoretical analysis,'' {\em Phys. Rev. Lett.}, vol.~109, p.~196802,
  2012.

\bibitem{Flores2013}
M.~Flores, E.~Cisternas, J.~Correa, and P.~Vargas, ``Moir\'e
patterns on stm
  images of graphite induced by rotations of surface and subsurface layer,''
  {\em Chemical Physics}, vol.~423, pp.~49--54, 2013.

\bibitem{yin14}
L.-J. Yin, J.-B. Qiao, W.-X. Wang, Z.-D. Chu, K.~F. Zhang, R.-F.
Dou, C.~L.
  Gao, J.-F. Jia, J.-C. Nie, and L.~He, ``Tuning structures and electronic
  spectra of graphene layers with tilt grain boundaries,'' {\em Phys. Rev. B},
  vol.~89, p.~205410, May 2014.

\bibitem{Burkov2011}
A.~A. Burkov and L.~Balents, ``Weyl semimetal in a topological
insulator
  multilayer,'' {\em Phys. Rev. Lett.}, vol.~107, p.~127205, Sep 2011.

\bibitem{Volovik2007}
G.~E. Volovik, ``Phase Transitions from Topology in Momentum
space'',
  vol.~718, pp.~31--73, 2007.

\bibitem{McClure1957}
J.~W. McClure, ``Band structure of graphite and de haas-van alphen
effect,''
  {\em Phys. Rev.}, vol.~108, pp.~612--618, Nov 1957.

\bibitem{HasanKane2010}
M.~Z. Hasan and C.~L. Kane, ``Colloquium,'' {\em Rev. Mod. Phys.},
vol.~82,
  pp.~3045--3067, Nov 2010.

\bibitem{Shoucheng2011}
X.-L. Qi and S.-C. Zhang, ``Topological insulators and
superconductors,'' {\em
  Rev. Mod. Phys.}, vol.~83, pp.~1057--1110, Oct 2011.

\bibitem{Ryu2002}
S.~Ryu and Y.~Hatsugai, ``Topological origin of zero-energy edge
states in
  particle-hole symmetric systems,'' {\em Phys. Rev. Lett.}, vol.~89,
  p.~077002, 2002.

\bibitem{Kopnin1991}
Y.~Kopnin and M.~Salomaa, ``Mutual friction in superfluid $^3$he:
effects of
  bound states in the vortex core,'' {\em Phys. Rev. B}, vol.~44, p.~9667,
  1991.

\bibitem{Volovik1994}
G.~Volovik, ``On fermi condensate: near the saddle point and
within the vortex
  core,'' {\em JETP Lett.}, vol.~59, pp.~830--835, 1994.

\bibitem{Volovik1999}
G.~Volovik, ``Fermion zero modes on vortices in chiral
superconductors,'' {\em
  JETP Lett.}, vol.~70, pp.~609--614, 1999.

\bibitem{Feng2012}
L.~Feng, X.~Lin, L.~Meng, J.-C. Nie, J.~Ni, and L.~He, ``Flat
bands near fermi
  level of topological line defects on graphite,'' {\em Appl. Phys. Lett.},
  vol.~101, p.~113113, 2012.

\bibitem{HeikkilaKopniVolovik2011}
T.~Heikkil\"a, N.~B. Kopnin, and G.~Volovik, ``Flat bands in
topological
  media,'' {\em JETP Lett.}, vol.~94, pp.~233--239, 2011.

\bibitem{kop13}
N.~B. Kopnin, M.~Ij\"as, A.~Harju, and T.~T. Heikkil\"a,
``High-temperature
  surface superconductivity in rhombohedral graphite,'' {\em Phys. Rev. B},
  vol.~87, p.~140503, 2013.

\bibitem{TangFu2014}
E.~Tang and L.~Fu, ``Strain-induced helical flat band and
interface
  superconductivity in topological crystalline insulators,'' {\em
  arXiv:1403.7523}, 2014.

\bibitem{Vozmedioano2013}
F.~de~Juan, J.~L. Ma\~nes, and M.~A.~H. Vozmediano, ``Gauge fields
from strain
  in graphene,'' {\em Phys. Rev. B}, vol.~87, p.~165131, Apr 2013.

\bibitem{Khodel1990}
V.~Khodel and V.~Shaginyan, ``Topological origin of zero-energy
edge states in
  particle-hole symmetric systems,'' {\em JETP Lett.}, vol.~51, p.~553, 1990.

\bibitem{kop11}
N.~B. Kopnin, T.~T. Heikkil\"a, and G.~E. Volovik,
``High-temperature surface
  superconductivity in topological flat-band systems,'' {\em Phys. Rev. B},
  vol.~83, p.~220503, 2011.

\bibitem{Kopaev1970}
Y.~Kopaev, ``Superconducting of alloyed semimetals,'' {\em JETP},
vol.~31,
  pp.~544--547, 1970.

\bibitem{KopaevRusinov1987}
Y.~Kopaev and A.~Rusinov, ``Enhancement of superconducting
critical temperature
  due to metal-semiconductor transition,'' {\em Phys. Lett. A}, vol.~121,
  pp.~300--304, 1987.

\bibitem{Dolgopolov2014}
A.~Shashkin, V.~Dolgopolov, J.~Clark, V.~Shaginyan, M.~Zverev, and
V.~Khodel,
  ``Merging of landau levels in a strongly-interacting two-dimensional electron
  system in silicon,'' {\em arXiv:1404.7465}, 2014.

\bibitem{Yudin2014}
D.~Yudin, D.~Hirschmeier, H.~Hafermann, O.~Eriksson, A.~I.
Lichtenstein, and
  M.~I. Katsnelson, ``Fermi condensation near van hove singularities within the
  hubbard model on the triangular lattice,'' {\em Phys. Rev. Lett.}, vol.~112,
  p.~070403, Feb 2014.

\bibitem{ant74}
K.~Antonowicz, ``Possible superconductivity at room temperature,''
{\em
  Nature}, vol.~247, pp.~358--360, 1974.

\bibitem{yakovjltp00}
Y.~Kopelevich, P.~Esquinazi, J.~Torres, and S.~Moehlecke,
``Ferromagnetic- and
  superconducting-like behavior of graphite,'' {\em J. Low Temp. Phys.},
  vol.~119, pp.~691--702, 2000.

\bibitem{silvaprl}
R.~R. da~Silva, J.~H.~S. Torres, and Y.~Kopelevich, ``Indication
of
  superconductivity at \rm{35~K} in graphite-sulfur composites,'' {\em Phys.
  Rev. Lett.}, vol.~87, pp.~147001--1--4, 2001.

\bibitem{yang01}
H.-P. Yang, H.-H. Wen, Z.-W. Zhao, and S.-L. Li, ``Possible
superconductivity
  at 37 k in graphite-sulfur composites,'' {\em Chin. Phys. Lett.}, vol.~18,
  pp.~1648--1650, 2001.

\bibitem{mol02}
S.~Moehlecke, P.~C. Ho, and M.~B. Maple, ``Coexistence of
superconductivity and
  ferromagnetism in the graphite-sulphur system,'' {\em Phil. Mag. B}, vol.~82,
  pp.~1335--1347, 2002.

\bibitem{fel09}
I.~Felner and Y.~Kopelevich, ``Magnetization measurement of a
possible
  high-temperature superconducting state in amorphous carbon doped with
  sulfur,'' {\em Phys. Rev. B}, vol.~79, p.~233409, 2009.

\bibitem{fel14}
I.~Felner, ``Superconductivity and unusual magnetic behavior in
amorphous
  carbon,'' {\em Materials Research Express}, vol.~1, p.~016001, 2014.

\bibitem{esqpip}
P.~Esquinazi, ``Graphite and its hidden superconductivity,'' {\em
Papers in
  Physics}, vol.~5, p.~050007, 2013.

\end{thebibliography}

\end{document}